\documentstyle[epsfig]{fbssuppl}

\title{Towards a Unified Description of the Baryon Spectrum
and the Baryon-Baryon Interaction within a Potential Model Scheme}
\author{A. Valcarce\instnr{1}, P. Gonz\'alez\instnr{2},
F. Fern\'andez\instnr{1}, V. Vento\instnr{2}}
\instlist{Grupo de F\'\i sica Nuclear, Universidad de Salamanca,
E-37008 Salamanca, Spain
\and Departamento de F\' \i sica Te\'orica e IFIC,
Universidad de Valencia-CSIC, E-46100 Burjassot, Valencia, Spain}

\begin{document}

\maketitle
\begin{abstract}
We study the low energy part of the nucleon and $\Delta$ spectra
by solving the Schr\"{o}dinger equation for the three-quark system in
the hyperspherical harmonic approach. The quark-quark
hamiltonian considered includes, besides the usual one-gluon
exchange, pion and sigma exchanges generated by the
chiral symmetry breaking. This quark-quark potential
reproduces, in a Resonating Group Method calculation,
the nucleon-nucleon
scattering phase shifts and the deuteron properties.
The baryonic spectrum
obtained is quite reasonable and the resulting
wave function is consistent
with the ansatz used in the two baryon system.
\end{abstract}

\section{Introduction}

Models have been widely used to study the properties of the
hadron spectrum due to the impossibility to solve
Quantum Chromodynamics (QCD) at the
current moment.
In particular, the quark potential models incorporate
the perturbative one-gluon exchange quark-quark ($qq$) potential
($V_{\rm OGE}$) derived from QCD as well as a parametrization
of some nonperturbative effects through a $qq$ confining
potential ($V_{\rm con}$) \cite{RUJ}. Such an {\em effective
theory} [the quark-gluon coupling constant is taken as an
effective one and the constituent quark masses ($m_q$) are parameters
fitted from the baryon magnetic moments] provides a reasonable
understanding of the baryon spectrum and the hadron static
properties \cite{YAO}.

The idea of an interquark potential has been also
used to study baryon-baryon interactions. More
explicitly, the repulsive core of the nucleon-nucleon (NN)
force has been shown to arise from the color-spin structure
of the $V_{\rm OGE}$ \cite{FAE}. Nevertheless, the same
scheme has been proven incapable of describing the long
range part of the NN interaction unless pion exchange
between quarks is introduced. From the basic theory,
the origin of the one-pion exchange potential ($V_{\rm OPE}$)
is associated with the spontaneous chiral symmetry breaking
of QCD.
Moreover, the inclusion of the
$qq$ sigma potential ($V_{\rm OSE}$) consistently with its
chiral partner (the pion) allows to reproduce the
intermediate range $NN$ interaction as well as the
deuteron properties \cite{FER}.

In this paper we examine the consistency of the two scenarios.
We determine a $qq$ interaction by studying two nucleon
properties and charge exchange reactions and proceed to
analyze the baryon spectrum for this interaction. As we
shall discuss the predictions for the baryon masses and
the baryon wave functions, it is not only a stringent
test of the potential but also a consistency test of the
formalism used to generate the NN interaction
[resonating group method (RGM)].

\section{The Quark-Quark Potential}

The starting point of our description is
a quark-quark interaction of the following form:

\begin{equation}
V_{\rm qq} (\vec{r}_{ij}) = V_{\rm con} (\vec{r}_{ij}) +
V_{\rm OGE} (\vec{r}_{ij})
+ V_{\rm OPE} (\vec{r}_{ij})  + V_{\rm OSE} (\vec{r}_{ij}) \, ,
\end{equation}

\noindent
where $\vec{r}_{ij}$ is the interquark distance.

The confinement is chosen to be linear as suggested by the
meson spectrum and lattice calculations:

\begin{equation}
V_{\rm con} (\vec{r}_{ij}) = - a_c \,
( \vec{\lambda}_i \cdot \vec{\lambda}_j ) \, r_{ij} \, ,
\end{equation}

\noindent
where $\lambda 's$ are the SU(3) color matrices.

$V_{\rm OGE}$, $V_{\rm OPE}$ and $V_{\rm OSE}$ have been derived in detail
elsewhere \cite{RUJ,FER} and we shall limit here to write
the final expressions:

\begin{equation}
V_{\rm OGE} ({\vec r}_{ij}) =
{1 \over 4} \, \alpha_s \, {\vec
\lambda}_i \cdot {\vec \lambda}_j
\Biggl \lbrace {1 \over r_{ij}} -
{1 \over {4 \, m^2_q}} \, \biggl [ 1 + {2 \over 3}
{\vec \sigma}_i \cdot {\vec
\sigma}_j \biggr ] \,\,
{{e^{-r_{ij}/r_0}} \over
{r_0^2 \,\,r_{ij}}}
- {1 \over {4 m^2_q \, r^3_{ij}}}
\, S_{ij} \Biggr \rbrace \, ,
\end{equation}

\noindent
where $\alpha_s$ is the effective quark-quark-gluon coupling constant,
$r_0$ is the range of a smeared $\delta$ function in order to avoid
and unbound spectrum \cite{BHA},
the $\sigma ' s$
stand for the spin Pauli matrices and $S_{ij}$ is the quark tensor operator
$S_{ij} = 3 (\vec{\sigma}_i \, . \, \hat{r}_{ij}) (\vec{\sigma}_j
\, . \, \hat{r}_{ij}) - \vec{\sigma}_i \, . \vec{\sigma}_j $.

\begin{eqnarray}
V_{\rm OPE} ({\vec r}_{ij}) & = & {1 \over 3}
\, \alpha_{ch} {\Lambda^2  \over \Lambda^2 -
m_\pi^2} \, m_\pi \, \Biggr\{ \left[ \,
Y (m_\pi \, r_{ij}) - { \Lambda^3
\over m_{\pi}^3} \, Y (\Lambda \,
r_{ij}) \right] {\vec \sigma}_i \cdot
{\vec \sigma}_j + \nonumber \\
 & & \left[ H( m_\pi \, r_{ij}) - {
\Lambda^3 \over m_\pi^3} \, H( \Lambda \,
r_{ij}) \right] S_{ij} \Biggr\} \,
{\vec \tau}_i \cdot {\vec \tau}_j \, ,
\end{eqnarray}

\begin{equation}
V_{\rm OSE} ({\vec r}_{ij}) = - \alpha_{ch} \,
{4 \, m_q^2 \over m_{\pi}^2}
{\Lambda^2 \over \Lambda^2 - m_{\sigma}^2}
\, m_{\sigma} \, \left[
Y (m_{\sigma} \, r_{ij})-
{\Lambda \over {m_{\sigma}}} \,
Y (\Lambda \, r_{ij}) \right] \, ,
\end{equation}

\noindent
where $m_\pi$ ($m_\sigma$) is the pion (sigma) mass,
$\alpha_{ch}$ is the chiral coupling constant (related to the
$\pi NN$ coupling constant), $\Lambda$ is a cutoff
parameter and Y(x), H(x) are the Yukawa functions defined as:

\begin{equation}
Y(x) \, = \, {e^{-x} \over x} \,\,\,\, , \,\,\,\,
H(x) \, = \, \Bigl( 1 + {3 \over x} + { 3 \over {x^2}} \Bigr) Y(x) \, ,
\end{equation}

With this interaction, the two baryon system has been studied in
the RGM framework assuming
for the spatial part of the
wave function of the quarks a harmonic
oscillator ground state,

\begin{equation}
\eta_{\rm os} (\vec{r}_i - \vec{R} \, ) =
\left( {1 \over {\pi b^2}} \right)^{3/4} e^{-(\vec{r}_i
- \vec{R}\,)^2/ 2b^2} \, ,
\end{equation}

\noindent
where the parameter $\vec{R}$
determines the position of the baryon and
$b$ is the harmonic oscillator constant.

\begin{figure}[t]
\vbox{
\vspace*{-2.75cm}
\centerline{\epsfig{file=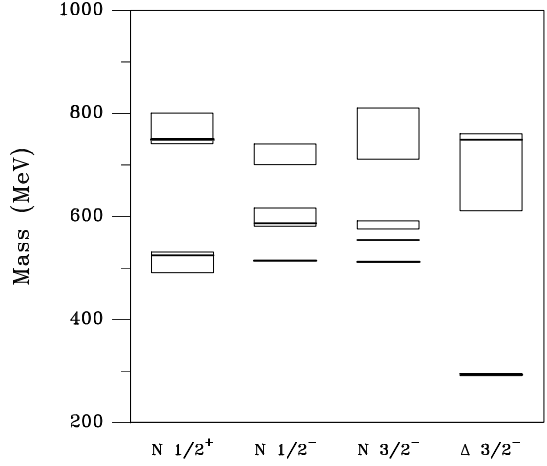,height=5.2in}}
\vspace*{-3.5cm}
\caption{Relative energy $N$ and $\Delta$ spectrum
up to 0.7 GeV excitation energy. The solid line corresponds
to the results of our model. The boxes represent the
experimental data with the corresponding uncertainties.}
\label{espe}
}
\end{figure}

\begin{table}[hbt]
\caption[]{Value of the potential parameters}
\label{param}
\begin{tabular}{|c|c|c|c|c|c|c|c|}
\hline
\tabstrut $m_q$ & $b$ & $\alpha_s$ & $\alpha_{ch}$ & $a_c$ &
$m_\pi$ & $m_\sigma$ & $\Lambda$ \\
MeV & fm &  &  & ${\rm MeV} \cdot {\rm fm}^{-1}$ & ${\rm fm}^{-1}$
& ${\rm fm}^{-1}$ & ${\rm fm}^{-1}$ \\
\hline
313 & 0.518 & 0.485 & 0.027 & 91.488 & 0.7 & 3.42 & 4.2 \\
\hline
\end{tabular}
\end{table}

Using the same set of parameters given in Table \ref{param}, the $NN$
scattering phase shifts, the static and electromagnetic
properties of the deuteron \cite{FER} and reactions that
take place with the excitation of the $\Delta$ resonance
\cite{FER2} are reasonably reproduced.

\section{Results}

To study the baryonic spectrum with the
potential just described (parameters
as in Table \ref{param})
we solve the Schr\"{o}dinger equation in
the hyperspherical harmonic
approach \cite{BAL}. The low energy $N$ and $\Delta$
spectrum obtained, a part of which is shown in Fig. \ref{espe},
is quite reasonable though some small discrepancy remains
concerning the relative energy position of the Roper
resonance [$N^* (1440)$] and the first negative parity
state excitation [$N^- (1535)$], as it is common in two body
potential models.

\begin{figure}[t]
\vbox{
\vspace*{-3.0cm}
\centerline{\epsfig{file=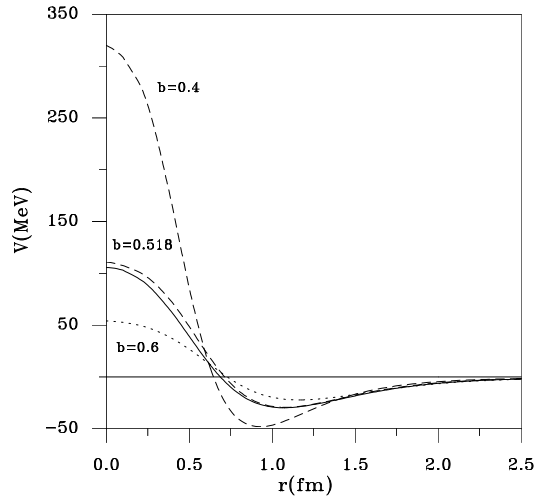,height=4.0in}}
\vspace*{-0.1cm}
\caption{Adiabatic $NN$ potential for different nucleon
wave functions for the channel (S,T)=(1,0).
The solid line corresponds to the solution
of the Schr\"odinger equation. The others to a gaussian
ansatz with the values of $b$ given in the figure.}
\label{pot}
}
\end{figure}

Regarding the wave function and in order to check the internal
RGM consistency, it makes sense to compare the $NN$ potential
one obtains from the wave function solution of the Schr\"odinger
equation, with the potential obtained from the ansatz RGM
wave function for different values of the parameter $b$. For the
sake of technical simplicity we do this in the Born-Oppeheimer
approximation. The results appear in Fig. \ref{pot}, where we can see
that it is precisely the value of $b$ that gives
the best overall fit to the $NN$ data with RGM, the one
which provides a better approximation. This confers to
$b$ a self-consistent character and solves the controversy
about its possible values.

Certainly, more work and further refinements are needed along
this direction. Meanwhile, the proposed
model points out the plausibility of a
unified description of the baryon structure and the
baryon-baryon interaction.

\begin{acknowledge}
This work has been partially funded by Direcci\'on
General de Investigaci\'on Cient\'{\i}fica y T\'ecnica
(DGICYT) under the Contract No. PB91-0119
\end{acknowledge}

\catcode`\@=11 \if@amssymbols%
\clearpage 
\else\relax\fi\catcode`\@=12

\SaveFinalPage
\end{document}